\documentclass[12pt,twoside]{article}

\usepackage{wrapfig}
\usepackage{graphicx}
\usepackage{cmp2e}

\title[Spectrum of the Hubbard model]{Evolution of the spectrum of the
Hubbard model with filling}

\author{A.~Sherman}

\address{Institute of Physics, University of Tartu, Riia 142, 51014 Tartu,
Estonia}

\begin{document}

\maketitle

\begin{abstract}
The diagram technique for the one-band Hubbard model is formulated for
the case of moderate to strong Hubbard repulsion. The expansion in
powers of the hopping constant is expressed in terms of site cumulants
of electron creation and annihilation operators. For Green's function
an equation of the Larkin type is derived and solved in a one-loop
approximation for the case of two dimensions and nearest-neighbor
hopping. With decreasing the electron concentration in addition to the
four bands observed at half-filling a narrow band arises near the Fermi
level. The dispersion of the new band, its bandwidth and the variation
with filling are close to those of the spin-polaron band in the $t$-$J$
model.
\pacs 71.10.Fd, 71.10.-w
\end{abstract}

\section{Introduction}
Systems with strong electron correlations which, in particular, cuprate
perovskites belong to are characterized by a Coulomb repulsion that is
comparable to or larger than hopping constants. One of the simplest
models for the description of such systems is the Hubbard model
\cite{Gutzwiller,Hubbard63,Kanamori} with the Hamiltonian
\begin{equation}\label{Hamiltonian}
H=\sum_{\bf nm\sigma}t_{\bf nm}a^\dagger_{\bf n\sigma}a_{\bf m\sigma} +
\frac{U}{2}\sum_{\bf n\sigma}n_{\bf n\sigma}n_{\bf n,-\sigma},
\end{equation}
where $t_{\bf nm}$ is the hopping constants, the operator
$a^\dagger_{\bf n\sigma}$ creates an electron on the site {\bf n} with
the spin projection $\sigma=\pm 1$, $U$ is the on-site Coulomb
repulsion and the electron number operator $n_{\bf
n\sigma}=a^\dagger_{\bf n\sigma}a_{\bf n\sigma}$. In the case of strong
electron correlations, $U\geq|t_{\bf nm}|$, it is reasonable to use a
perturbation expansion in powers of the hopping constants for
investigating Hamiltonian (\ref{Hamiltonian}). Apparently the first
such expansion was considered in reference~\cite{Hubbard66}. The
further development of this approach was given in
references~\cite{Westwanski,Stasyuk,Zaitsev,Izyumov,Ovchinnikov} where
the diagram technique for Hubbard operators was developed and used for
investigating the Mott transition, the magnetic phase diagram and the
superconducting transition in the Hubbard model.

The rules of the diagram technique for Hubbard operators are rather
intricate. Besides, these rules and the graphical representation of the
expansion vary depending on the choice of the operator precedence. The
diagram technique suggested in
references~\cite{Vladimir,Metzner,Moskalenko,Pairault} is free from
these defects. In this approach the power expansion for Green's
function of electron operators $a_{\bf n\sigma}$ and $a^\dagger_{\bf
n\sigma}$ is considered and the terms of the expansion are expressed as
site cumulants of these operators. Based on this diagram technique the
equations of the Larkin type \cite{Larkin} for Green's function were
derived \cite{Vladimir,Moskalenko,Pairault}.

However, the application of this approach runs into problems. In
particular, at half-filling the spectral weight obtained after a
resummation of diagrams appears to be negative near frequencies
$\omega_d=\pm U/2$ \cite{Pairault}. This drawback is connected with
divergencies in cumulants at these frequencies \cite{Sherman}. As can
be seen from formulas given below, all higher-order cumulants have such
divergencies at $\omega_d$ with sign-changing residues, which are
expected to compensate the negative spectral weight in the entire
series. On the other hand, at frequencies neighboring to $\omega_d$
cumulants are regular. If a selected subset of diagrams is expected to
give a correct estimate of the entire series for these frequencies the
values at $\omega_d$ can be corrected by an interpolation using results
for the regular regions. This procedure was applied in
reference~\cite{Sherman} for the case of two dimensions, half-filling,
nearest-neighbor hopping and with the use of the one-loop
approximation. The spectrum was shown to consist of four bands. These
band structure and the calculated shapes of the electron spectral
function are close to those obtained in the Monte-Carlo
\cite{Moreo,Preuss,Grober}, cluster perturbation \cite{Dahnken} and the
two-particle self-consistent \cite{Tremblay} calculations. The
four-band structure of the spectrum of the Hubbard model was also
considered in reference~\cite{Shvaika}.

In the approach of reference~\cite{Sherman} the mentioned four-band
structure of the spectrum at half-filling has its origin in the regions
of large damping which separate the low- and high-frequency bands. In
the major part of the Brillouin zone both these types of the bands owe
their origin to the same terms of the irreducible part, i.e.\ to the
same interaction processes. This brings up the questions: How this
spectrum is changed with filling and how the quasiparticle peak, which
determines the photoemission leading edge, arises in the spectrum? As
expected, this peak differs in nature from other spectral features. In
the present work it is shown that on certain deviation from
half-filling an additional narrow band arises near the Fermi level on
the background of the four above-mentioned bands. The dispersion of the
new band, its bandwidth and the variation with filling are close to
those of the spin-polaron band in the $t$-$J$ model
\cite{Plakida,Sherman94}.

In the following section the perturbation expansion for the electron
Green's function is formulated in the form convenient for calculations
and Larkin's equation is derived. In Sec.~III the equations of the
previous section are used for calculating the spectral function and the
obtained results are discussed. Concluding remarks are presented in
Sec.~IV.

\section{Diagram technique}
We consider Green's function
\begin{equation}\label{GF}
G({\bf n'\tau',n\tau})=\langle{\cal T}\bar{a}_{\bf n'\sigma}(\tau')
a_{\bf n\sigma}(\tau)\rangle,
\end{equation}
where the angular brackets denote the statistical averaging with the
Hamiltonian ${\cal H}=H-\mu\sum_{\bf n\sigma}n_{\bf n\sigma}$, $\mu$ is
the chemical potential, ${\cal T}$ is the time-ordering operator which
arranges other operators from right to left in ascending order of times
$\tau$, $a_{\bf n\sigma}(\tau)=\exp({\cal H}\tau)a_{\bf
n\sigma}\exp(-{\cal H}\tau)$ and $\bar{a}_{\bf
n\sigma}(\tau)=\exp({\cal H}\tau)a^\dagger_{\bf n\sigma}\exp(-{\cal
H}\tau)$. Choosing
\begin{equation}
H_0=\frac{U}{2}\sum_{\bf n\sigma}n_{\bf n\sigma}n_{\bf
 n,-\sigma}-\mu\sum_{\bf n\sigma}n_{\bf n\sigma}\;\;{\rm and}\;\;
H_1=\sum_{\bf nm\sigma}t_{\bf nm}a^\dagger_{\bf n\sigma}a_{\bf
 m\sigma}\label{division}
\end{equation}
as the unperturbed Hamiltonian and the perturbation, respectively, and
using the known expansion \cite{Abrikosov} for the evolution operator
we get
\begin{eqnarray}
G({\bf n'\tau',n\tau})&=&\sum_{k=0}^\infty \frac{(-1)^k}{k!}
 \int\!\ldots\!\int_0^\beta {\rm d}\tau_1\ldots {\rm d}\tau_k
 \sum_{{\bf n}_1{\bf n}^\prime_1\sigma_1}\!\!\ldots\!\!
 \sum_{{\bf n}_k{\bf n}^\prime_k\sigma_k} t_{{\bf n}_1{\bf n}^\prime_1}
 \ldots t_{{\bf n}_k{\bf n}^\prime_k}\nonumber\\
&\times&\langle{\cal T}\bar{a}_{\bf n'\sigma}(\tau') a_{\bf
 n\sigma}(\tau) \bar{a}_{{\bf n}^\prime_1\sigma_1}(\tau_1) a_{{\bf
 n}_1\sigma_1}(\tau_1)\ldots
 \bar{a}_{{\bf n}^\prime_k\sigma_k}(\tau_k)
 a_{{\bf n}_k\sigma_k}(\tau_k)\rangle_{\rm 0c}, \label{series}
\end{eqnarray}
where $\beta=T^{-1}$ is the inverse temperature, the subscript ``0''
near the angular bracket indicates that the averaging and time
dependencies of operators are determined with the Hamiltonian $H_0$.
The subscript ``c'' indicates that terms which split into two or more
disconnected averages have to be dropped out.

The Hamiltonian $H_0$ is diagonal in the site representation. Therefore
the average in the right-hand side of equation~(\ref{series}) splits
into averages belonging to separate lattice sites. To be nonzero these
latter averages have to contain equal number of creation and
annihilation operators. Let us consider the average which appears in
the first order: $\langle{\cal T}\bar{a}_{\bf n'\sigma}(\tau') a_{\bf
 n\sigma}(\tau) \bar{a}_{{\bf n}^\prime_1\sigma_1}(\tau_1) a_{{\bf
 n}_1\sigma_1}(\tau_1)\rangle$ (for short here and below subscripts
``0'' and ``c'' are omitted). For this average to be nonvanishing
operators have to belong either to the same site or to two different
sites,
\begin{eqnarray*}
&&\langle{\cal T}\bar{a}_{\bf n'\sigma}(\tau') a_{\bf
 n\sigma}(\tau) \bar{a}_{{\bf n}^\prime_1\sigma_1}(\tau_1^\prime)
 a_{{\bf n}_1\sigma_1}(\tau_1)\rangle\\
&&\quad=\langle{\cal T}\bar{a}_{\bf n\sigma}(\tau')
 a_{\bf n\sigma}(\tau) \bar{a}_{{\bf n}\sigma_1}(\tau_1^\prime)
 a_{{\bf  n}\sigma_1}(\tau_1)\rangle \delta_{\bf nn'}
 \delta_{{\bf nn}_1^\prime} \delta_{{\bf nn}_1}\\
&&\quad+\langle{\cal T}\bar{a}_{\bf n\sigma}(\tau') a_{\bf
 n\sigma}(\tau)\rangle \langle{\cal T}\bar{a}_{{\bf n}_1\sigma_1}
 (\tau_1^\prime) a_{{\bf n}_1\sigma_1}(\tau_1)\rangle\delta_{\bf nn'}
 \delta_{{\bf n}_1{\bf n}_1^\prime} (1-\delta_{{\bf nn}_1})\\
&&\quad-\langle{\cal T}\bar{a}_{{\bf n}_1\sigma}(\tau') a_{{\bf
 n}_1\sigma_1}(\tau_1)\rangle \langle{\cal T}\bar{a}_{{\bf n}\sigma_1}
 (\tau_1^\prime) a_{\bf n\sigma}(\tau)\rangle\delta_{{\bf n'n}_1}
 \delta_{{\bf nn}_1^\prime} (1-\delta_{{\bf nn}_1}).
\end{eqnarray*}
The multiplier $1-\delta_{{\bf nn}_1}$ ensures that ${\bf n}\neq{\bf
n}_1$ in the two last terms (the case ${\bf n}={\bf n}_1$ is taken into
account by the first term in the right-hand side). After the
rearrangement of the terms we find
\begin{eqnarray*}
&&\langle{\cal T}\bar{a}_{\bf n'\sigma}(\tau') a_{\bf
 n\sigma}(\tau) \bar{a}_{{\bf n}^\prime_1\sigma_1}(\tau_1^\prime)
 a_{{\bf n}_1\sigma_1}(\tau_1)\rangle=
 K_2(\tau'\sigma,\tau\sigma,\tau'_1\sigma_1,\tau_1\sigma_1)
 \delta_{\bf nn'} \delta_{{\bf nn}_1^\prime} \delta_{{\bf nn}_1}\\
&&\quad\quad+K_1(\tau'\sigma,\tau\sigma)K_1(\tau'_1\sigma_1,\tau_1
 \sigma_1)\delta_{\bf nn'} \delta_{{\bf n}_1{\bf n}'_1}
 -K_1(\tau'\sigma,\tau_1\sigma_1)K_1(\tau'_1\sigma_1,\tau\sigma)
 \delta_{{\bf n'n}_1} \delta_{{\bf nn}'_1},
\end{eqnarray*}
where
\begin{eqnarray}
&&K_1(\tau'\sigma',\tau\sigma)=\langle{\cal T}\bar{a}_\sigma(\tau')
 a_\sigma(\tau)\rangle \delta_{\sigma\sigma'}, \nonumber\\
&&K_2(\tau'\sigma,\tau\sigma,\tau'_1\sigma_1,\tau_1\sigma_1)=
 \langle{\cal T}\bar{a}_\sigma(\tau') a_\sigma(\tau)
 \bar{a}_{\sigma_1}(\tau'_1) a_{\sigma_1}(\tau_1)\rangle
 \label{cumulants}\\
&&\quad\quad
 -K_1(\tau'\sigma,\tau\sigma)K_1(\tau'_1\sigma_1,\tau_1\sigma_1)
 +K_1(\tau'\sigma,\tau_1\sigma_1)K_1(\tau'_1\sigma_1,\tau\sigma)
 \nonumber
\end{eqnarray}
are cumulants \cite{Kubo} of the first and second order, respectively.
All operators in cumulants~(\ref{cumulants}) belong to the same lattice
site. Due to the translational symmetry of $H_0$ in
equation~(\ref{division}) the cumulants do not depend on the site index
which is therefore omitted in equation~(\ref{cumulants}).

\begin{wrapfigure}{i}{0.6\textwidth}
\centerline{\includegraphics[width=0.56\textwidth]{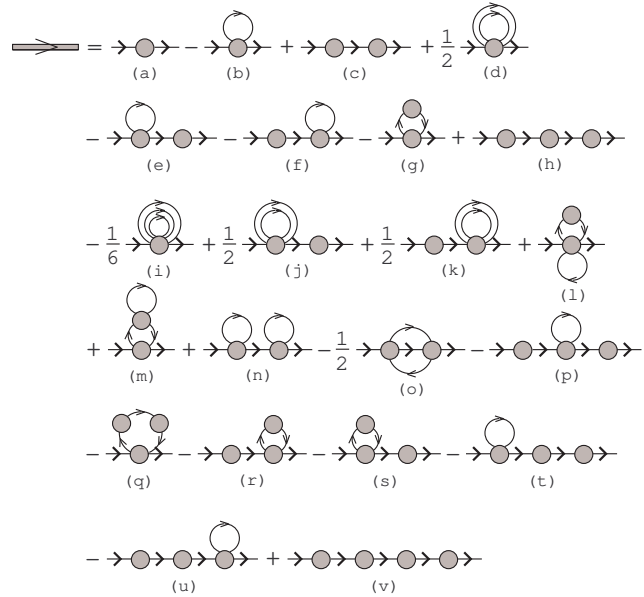}}
\caption{Diagrams of the first four orders of expansion
(\protect\ref{series}).}\label{Fig_i}
\end{wrapfigure}
Averages which appear in higher-order terms of
equati\-on~(\ref{series}) can be transformed in the same manner. The
average in the $k$-th order term which contains $k+1$ creation and
annihilation operators is represented by the sum of all possible
products of cumulants with the sum of orders equal to $k+1$. All
possible distributions of operators between the cumulants in these
products have to be taken into account. The sign of a term in this sum
is determined by the number of permutations of fermion operators, which
bring the sequence of operators in the initial average to that in the
term. Actually the above statements determine rules of the diagram
technique. Additionally we have to take into account the presence of
topologically equivalent terms -- terms which differ only by
permutation of operators $H_1(\tau_i)$ in equation~(\ref{series}).
Since these terms are equal, in the expansion only one of them can be
taken into account with the prefactor $\nu=j/k!$ where $j$ is the
number of topologically equivalent terms. Following
reference~\cite{Metzner} in diagrams we denote a cumulant by a circle
and the hopping constant $t_{\bf nn'}$ by a line directed from ${\bf
n'}$ to ${\bf n}$. The external operators $\bar{a}_{\bf
n'\sigma}(\tau')$ and $a_{\bf n\sigma}(\tau)$ are denoted by directed
lines leaving from and entering into a cumulant. The order of a
cumulant is equal to a number of incoming or outgoing lines. Summations
and integrations over the internal indices ${\bf n}_i$, ${\bf n}'_i$,
$\sigma_i$ and $\tau_i$ are carried out. Since site indices of
operators included in a cumulant coincide, some site summations
disappear. Also some summations over $\sigma_i$ get lost, because in
any cumulant spin indices of creation and annihilation operators have
to match. Taking into account the multiplier $(-1)^k$ in
equation~(\ref{series}), the sign of the diagram is equal to $(-1)^l$
where $l$ is the number of loops formed by hopping lines.
Figure~\ref{Fig_i} demonstrates connected diagrams of the first four
orders of the power expansion~(\ref{series}) with their signs and
prefactors. Here the thick line with arrow in the left-hand side of the
equation is the total Green's function. Notice that if we set $t_{\bf
nn}=0$, contributions of the diagrams (b), (d)--(f), (i)--(n), (p),
(t), and (u) vanish. However, below a renormalized hopping parameter
will be introduced which is nonzero for coinciding site indices and
therefore the mentioned diagrams are retained in figure~\ref{Fig_i}.

All diagrams can be separated into two categories -- those which can be
divided into two parts by cutting some hopping line and those which
cannot be divided in this way \cite{Izyumov,Larkin}. These latter
diagrams are referred to as irreducible diagrams. In figure~\ref{Fig_i}
the diagrams (c), (e), (f), (h), (j), (k), (n), (p), and (r)--(v)
belong to the first category, while the others are from the second
category. The former diagrams can be constructed from the irreducible
diagrams by connecting them with hopping lines. Notice that a prefactor
$\nu$ of some composite diagram is the product of prefactors of its
irreducible parts [cf., e.g., the irreducible diagrams (a), (d) and the
composed diagrams (j), (k) in figure~\ref{Fig_i}]. If we denote the sum
of all irreducible diagrams as $K({\bf n'}\tau',{\bf n}\tau)$, the
equation for Green's function can be written as
\begin{equation}
G({\bf n'}\tau',{\bf n}\tau)=K({\bf n'}\tau',{\bf n}\tau)+
 \sum_{{\bf n}_1{\bf n}'_1} \int_0^\beta {\rm d}\tau_1 K({\bf
 n'}\tau',{\bf n}_1\tau_1)
 t_{{\bf n}_1{\bf n}'_1} G({\bf n}'_1\tau_1,{\bf n}\tau).
 \label{Larkin}
\end{equation}
This equation has the form of the Larkin equation \cite{Larkin}.

The partial summation can be carried out in the hopping lines of
cumulants by inserting the irreducible diagrams into these lines. In
doing so the hopping constant $t_{\bf nn'}$ in the respective formulas
is substituted by
\begin{equation}
\Theta({\bf n\tau,n'\tau'})=t_{\bf nn'}\delta(\tau-\tau')+\sum_{{\bf
 n}_1{\bf n}'_1}t_{{\bf nn}'_1}\int_0^\beta{\rm d}\tau_1
 K({\bf n}'_1\tau,{\bf n}_1\tau_1)
 \Theta({\bf n}_1\tau_1,{\bf n}'\tau'). \label{hopping}
\end{equation}
For the diagram (b) in figure~\ref{Fig_i} inserting irreducible
diagrams into the hopping line leads to the diagrams (g), (m), and (q).

Due to the translation and time invariance of the problem the
quantities in equations~(\ref{Larkin}) and~(\ref{hopping}) depend only
on differences of site indices and times. The use of the Fourier
transformation
\[G({\bf k},{\rm i}\omega_l)=\sum_{\bf n'}{\rm e}^{{\rm i}{\bf
k(n'-n)}}\int_0^\beta{\rm d}\tau'{\rm e}^{{\rm
i}\omega_l(\tau'-\tau)}G({\bf n}'\tau',{\bf n}\tau),\] where
$\omega_l=(2l+1)\pi T$ is the Matsubara frequency, simplifies
significantly these equations:
\begin{eqnarray}
G({\bf k},{\rm i}\omega_l)&=&\frac{K({\bf k},{\rm i}\omega_l)}{1-t_{\bf
k}K({\bf k},{\rm i}\omega_l)},\label{tLarkin}\\
\Theta({\bf k},{\rm i}\omega_l)&=&\frac{t_{\bf k}}{1-t_{\bf k}K({\bf
 k},{\rm i}\omega_l)}
 =t_{\bf k}+t_{\bf k}^2G({\bf k},{\rm i}\omega_l).\label{thopping}
\end{eqnarray}

In the approximation used below the total collection of irreducible
diagrams $K({\bf k},{\rm i}\omega_l)$ is substituted by the sum of the
two simplest diagrams (a) and (b) appearing in the first two orders of
the perturbation theory. Due to the form of the latter diagram this
approximation is referred to as the one-loop approximation. In the
diagram (b) the hopping line is renormalized in accordance with
equation~(\ref{thopping}). Thus,
\begin{equation}
K({\rm i}\omega_l)=K_1({\rm i}\omega_l)
 -T\sum_{l_1\sigma_1}K_2({\rm i}\omega_l\sigma,{\rm i}\omega_{l_1}\sigma_1,
 {\rm i}\omega_{l_1}\sigma_1)\frac{1}{N}\sum_{\bf k}t^2_{\bf k}G({\bf
 k},{\rm i}\omega_{l_1}),\label{approx}
\end{equation}
where $K_1({\rm i}\omega_l)$ and
\begin{eqnarray*}
&&K_2({\rm i}\omega_{l'}\sigma,{\rm i}\omega_l\sigma,{\rm i}
 \omega_{l'_1}\sigma_1,{\rm i}\omega_{l_1}\sigma_1)\\
&&\quad=\int\!\!\!\!\int\!\!\!\!\int\!\!\!\!\int_0^\beta {\rm d}\tau'
 {\rm d}\tau{\rm d}\tau'_1{\rm d}\tau_1{\rm e}^{-{\rm i}\omega_{l'}\tau'
 +{\rm i}\omega_l\tau-{\rm i}\omega_{l'_1}
 \tau'_1+{\rm i}\omega_{l_1}\tau_1}K_2(\tau'\sigma,\tau\sigma,\tau'_1\sigma_1,
 \tau_1\sigma_1)\\[1ex]
&&\quad=\beta\delta_{l+l_1,l'+l'_1}K_2({\rm i}\omega_l\sigma,{\rm
 i}\omega_{l'_1}\sigma_1,{\rm i}\omega_{l_1}\sigma_1)
\end{eqnarray*}
are the Fourier transforms of cumulants~(\ref{cumulants}), $N$ is the
number of sites and we set $\sum_{\bf k}t_{\bf k}=0$. Notice that in
this approximation $K$ does not depend on momentum.

Now we need to calculate the cumulants in equation~(\ref{approx}). To
do this it is convenient to introduce the Hubbard operators
$X^{ij}_{\bf n}=|i{\bf n}\rangle\langle j{\bf n}|$ where $|i{\bf
n}\rangle$ are eigenvectors of site Hamiltonians forming $H_0$,
equation~(\ref{division}). For each site there are four such states:
the empty state $|0{\bf n}\rangle$ with the energy $E_0=0$, the two
degenerate singly occupied states $|\sigma{\bf n}\rangle$ with the
energy $E_1=-\mu$ and the doubly occupied state $|2{\bf n}\rangle$ with
the energy $E_2=U-2\mu$. The Hubbard operators are connected by the
relations
\begin{equation}\label{Hubbard}
a_{\bf n\sigma}=X^{0\sigma}_{\bf n}+\sigma X^{-\sigma,2}_{\bf n},\quad
a^\dagger_{\bf n\sigma}=X^{\sigma 0}_{\bf n}+\sigma X^{2,-\sigma}_{\bf
n}
\end{equation}
with the creation and annihilation operators. The commutation relations
for the Hubbard operators are easily derived from their definition.
Using equation~(\ref{Hubbard}) the first cumulant in
equation~(\ref{cumulants}) can be computed straightforwardly:
\begin{equation}\label{firstc}
K_1({\rm i}\omega_l)=\frac{1}{Z_0}\left(\frac{{\rm e}^{-\beta
E_\sigma}+{\rm e}^{-\beta E_0}}{{\rm i}\omega_l-E_{\sigma0}}+\frac{{\rm
e}^{-\beta E_2}+{\rm e}^{-\beta E_\sigma}}{{\rm
i}\omega_l-E_{2\sigma}}\right),
\end{equation}
where $Z_0={\rm e}^{-\beta E_0}+2{\rm e}^{-\beta E_\sigma}+{\rm
e}^{-\beta E_2}$ is the partition function, $E_{ij}=E_i-E_j$. As
indicated in \cite{Izyumov,Vladimir,Pairault}, if $K({\bf k},{\rm
i}\omega_l)$ is approximated by this cumulant the result corresponds to
the Hubbard-I approximation \cite{Hubbard63}.

To calculate $K_2$ it is convenient to use Wick's theorem for Hubbard
operators \cite{Westwanski,Stasyuk,Zaitsev,Izyumov}:
\begin{eqnarray}
&&\langle{\cal T}X_{\alpha_1}(\tau_1)\ldots X_{\alpha_i}(\tau_i)
 X_\alpha(\tau)X_{\alpha_{i+1}}(\tau_{i+1})\ldots
 X_{\alpha_n}(\tau_n)\rangle\nonumber\\
&&\quad\quad=\sum_{k=1}^n (-1)^{P_k} g_\alpha(\tau-\tau_k)
 \langle{\cal T}X_{\alpha_1}(\tau_1)\ldots
 [X_{\alpha_k},X_\alpha]_\pm(\tau_k)\ldots
 X_{\alpha_n}(\tau_n)\rangle, \label{Wick}
\end{eqnarray}
where the averaging and time dependencies of operators are determined
by the Hamiltonian $H_0$, $\alpha$ is the index combining the state and
site indices of the Hubbard operator. If $X_\alpha$ is a fermion
operator ($X^{0\sigma}$, $X^{\sigma2}$ and their conjugates), $P_k$ is
the number of permutation with other fermion operators which is
necessary to transfer the operator $X_\alpha$ from its position in the
left-hand side of equation~(\ref{Wick}) to the position in the
right-hand side. In this case
\begin{equation}\label{fermion}
g_\alpha(\tau)=\frac{{\rm e}^{E_{ij}\tau}}{{\rm e}^{\beta
E_{ij}}+1}\left\{
\begin{array}{rl}-1, & \tau>0, \\ {\rm e}^{\beta E_{ij}}, & \tau<0,
\end{array} \right.
\end{equation}
where $i$ and $j$ are the state indices of $X_\alpha$. If $X_\alpha$ is
a boson operator ($X^{00}$, $X^{22}$, $X^{\sigma\sigma'}$, $X^{02}$,
and $X^{20}$), $P_k=0$ and
\begin{equation}\label{boson}
g_\alpha(\tau)=\frac{{\rm e}^{E_{ij}\tau}}{{\rm e}^{\beta
E_{ij}}-1}\left\{
\begin{array}{rl}1, & \tau>0, \\ {\rm e}^{\beta E_{ij}}, & \tau<0.
\end{array} \right.
\end{equation}
In equation~(\ref{Wick}), $[X_{\alpha_k},X_\alpha]_\pm$ denotes an
anticommutator when both operators are of fermion type and a commutator
in other cases.

The substitution of equation~(\ref{Hubbard}) in $\langle{\cal T}
\bar{a}_\sigma (\tau')a_\sigma(\tau)\bar{a}_{\sigma_1}(\tau'_1)
a_{\sigma_1}(\tau_1)\rangle$ in $K_2$, equation~(\ref{cumulants}),
leads to six nonvanishing averages of Hubbard operators (such averages
are nonzero if the numbers of the operators $X^{0\sigma}$ and
$X^{\sigma 0}$ coincide in them, and the same is true for the pair
$X^{\sigma 2}$ and $X^{2\sigma}$). Applying Wick's theorem (\ref{Wick})
to these averages the number of operators in them is sequentially
decreased until only time-independent operators are left. For $H_0$ in
equation~(\ref{division}) these are $X^{00}$, $X^{\sigma\sigma'}$, and
$X^{22}$. Their averages are easily calculated. As a result, after some
algebra we find
\begin{eqnarray}
&&\sum_{\sigma_1}K_2({\rm i}\omega_l\sigma,{\rm i}\omega_{l_1}\sigma_1,
 {\rm i}\omega_{l_1}\sigma_1)=\nonumber\\
&&\quad-Z_0^{-1}U \Bigl\{{\rm e}^{-\beta E_0}g_{0\sigma}
 ({\rm i}\omega_l)g_{0\sigma}({\rm i}\omega_{l_1})g_{02}({\rm i}\omega_l+
 {\rm i}\omega_{l_1})\Bigl[g_{0\sigma}({\rm i}\omega_l)+g_{0\sigma}
 ({\rm i}\omega_{l_1})\Bigr] \nonumber\\
&&\quad\quad+{\rm e}^{-\beta E_2}g_{\sigma2}({\rm
 i}\omega_l)g_{\sigma2}
 ({\rm i}\omega_{l_1})g_{02}({\rm i}\omega_l+{\rm i}\omega_{l_1})
 \Bigl[g_{\sigma2}({\rm i}\omega_l)+g_{\sigma2}({\rm i}\omega_{l_1})
 \Bigr]\nonumber\\
&&\quad\quad+{\rm e}^{-\beta E_1}\Bigl[g_{0\sigma}({\rm
 i}\omega_l)g_{\sigma2}({\rm i}\omega_l)
 \Bigl(g_{0\sigma}({\rm i}\omega_{l_1})-g_{\sigma2}
 ({\rm i}\omega_{l_1})\Bigr)^2\nonumber\\
&&\quad\quad\quad
 +g_{0\sigma}({\rm i}\omega_{l_1})g_{\sigma2}({\rm i}\omega_{l_1})
 \Bigl(g_{0\sigma}^2({\rm i}\omega_l)+g_{\sigma2}^2({\rm i}\omega_l)
 \Bigr)\Bigr]\Bigr\}\nonumber\\
&&\quad-Z_0^{-2}
 U^2\beta\delta_{ll_1}\Bigl({\rm e}^{-\beta(E_0+E_2)}
 +2{\rm e}^{-\beta(E_0+E_1)}+3{\rm e}^{-2\beta E_1}+
 2{\rm e}^{-\beta(E_1+E_2)}\Bigr)
 \nonumber\\
&&\quad\quad\times g_{0\sigma}^2({\rm i}\omega_l)
 g_{\sigma2}^2(i\omega_l) \nonumber\\
&&\quad+Z_0^{-2}U^2\beta
 \Bigl(2{\rm e}^{-\beta(E_0+E_2)}+{\rm e}^{-\beta(E_0+E_1)}
 +{\rm e}^{-\beta(E_1+E_2)}\Bigr)\nonumber\\
&&\quad\quad\times g_{0\sigma}({\rm i}\omega_l)g_{\sigma2}
 ({\rm i}\omega_l)g_{0\sigma}({\rm i}\omega_{l_1})g_{\sigma2}({\rm
 i}\omega_{l_1}),
 \label{secondc}
\end{eqnarray}
where $g_{ij}({\rm i}\omega_l)=({\rm i}\omega_l+E_{ij})^{-1}$ is the
Fourier transform of functions~(\ref{fermion}) and~(\ref{boson}).

Equations~(\ref{firstc}) and~(\ref{secondc}) can be significantly
simplified for the case of principal interest $U\gg T$. In this case if
$\mu$ satisfies the conditions
\begin{equation}\label{condition}
\lambda<\mu<U-\lambda,
\end{equation}
$\lambda\gg T$, the exponent $\exp(-\beta E_1)$ is much larger than
$\exp(-\beta E_0)$ and $\exp(-\beta E_2)$. By passing to real
frequencies we can ascertain that terms in (\ref{secondc}) with the two
latter multipliers contain the same peculiarities as other terms.
Therefore terms with these multipliers can be omitted in the equations
and we get
\begin{eqnarray}
&&K_1({\rm i}\omega_l)=\frac{{\rm i}\omega_l+\mu-U/2}{({\rm
 i}\omega_l+\mu)({\rm i}\omega_l+\mu-U)},\nonumber\\
&&\sum_{\sigma_1}K_2({\rm i}\omega_l\sigma,{\rm i}\omega_{l_1}\sigma_1,
 {\rm i}\omega_{l_1}\sigma_1)
 =-\frac{1}{2}Ug_{0\sigma}({\rm i}\omega_l)g_{\sigma 2}({\rm i}\omega_l)
 \Bigl[g_{0\sigma}^2({\rm i}\omega_{l_1})+g_{\sigma 2}^2({\rm i}
 \omega_{l_1})\Bigr]\label{simplified}\\
&&\quad\quad-\frac{1}{2}Ug_{0\sigma}({\rm i}\omega_{l_1})g_{\sigma
 2}({\rm i}\omega_{l_1})\Bigl[g_{0\sigma}({\rm i}\omega_l)-
 g_{\sigma 2}({\rm i}\omega_l)\Bigr]^2-\frac{3}{4}U^2\beta
 \delta_{ll_{1}}g_{0\sigma}^2({\rm i}\omega_l)
 g_{\sigma 2}^2({\rm i}\omega_l).\nonumber
\end{eqnarray}

Let us turn to real frequencies by substituting ${\rm i}\omega_l$ with
$z=\omega+{\rm i}\eta$ where $\eta$ is a small positive constant which
affords an artificial broadening. Results given in the next section
were calculated with $G({\bf k},z)$ in the right-hand side of
equation~(\ref{approx}) taken from the Hubbard-I approximation. As
mentioned, this Green's function is obtained if $K({\bf k}\omega)$ in
equation~(\ref{tLarkin}) is approximated by $K_1$ from
equation~(\ref{simplified}) which gives
\begin{eqnarray}
&&G({\bf k},z)=\frac{1}{2}\left(1+\frac{t_{\bf k}}{\sqrt{U^2+t_{\bf
 k}^2}}\right)\frac{1}{z-\varepsilon_{1,\bf k}}
 +\frac{1}{2}\left( 1-\frac{t_{\bf k}}{\sqrt{U^2+t_{\bf k}^2}}\right)
 \frac{1}{z-\varepsilon_{2,\bf k}},\nonumber\\
&&\label{HubbardI}\\
&&\varepsilon_{1,\bf k}=\frac{1}{2}\left(U+t_{\bf k}+\sqrt{U^2+t_{\bf
 k}^2}\right)-\mu,\quad
 \varepsilon_{2,\bf k}=\frac{1}{2}\left(U+t_{\bf k}-
 \sqrt{U^2+t_{\bf k}^2}\right)-\mu.\nonumber
\end{eqnarray}
Below the two-dimensional square lattice is considered. It is supposed
that only the hopping constants between nearest neighbor sites $t$ are
nonzero which gives $t_{\bf k}=2t[\cos(k_x)+\cos(k_y)]$ where the
intersite distance is taken as the unit of length. Due to the
electron-hole symmetry of Hamiltonian~(\ref{Hamiltonian}) the
consideration can be restricted to the range of the chemical potentials
$\mu\leq \frac{U}{2}$.

\section{Spectral function}
\begin{figure}
\centerline{\includegraphics[width=0.48\textwidth]{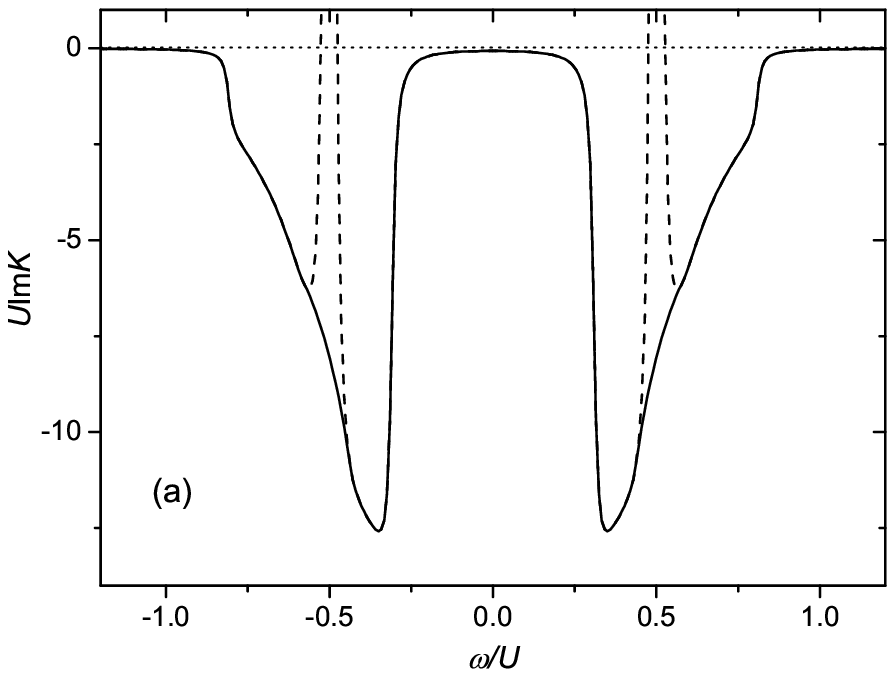}
\raisebox{-0.4ex}{\includegraphics[width=0.497\textwidth]{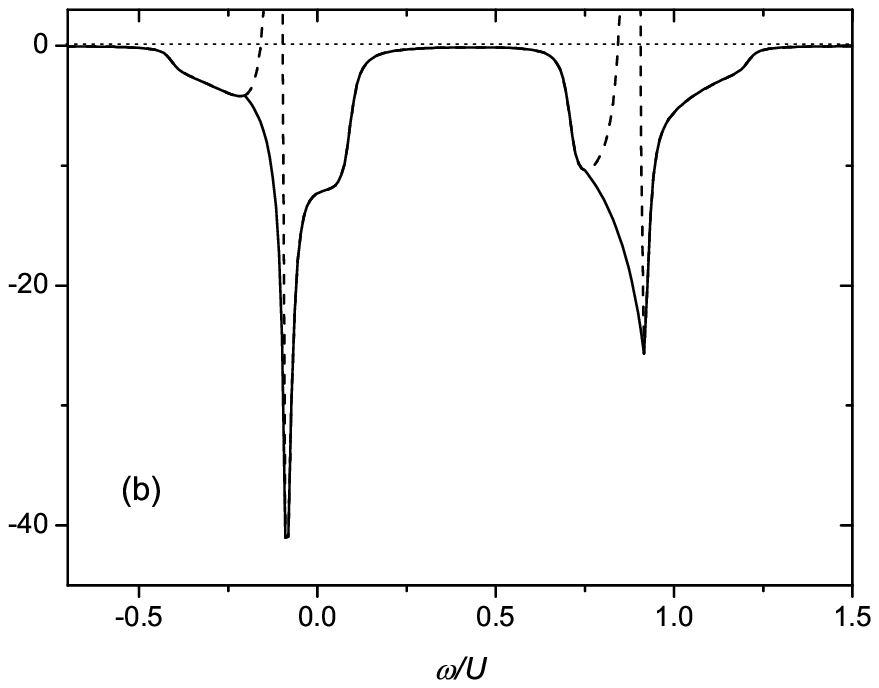}}}
\caption{The imaginary part of $K(\omega)$ calculated using
equations~(\protect\ref{approx}), (\ref{simplified})
and~(\protect\ref{HubbardI}) for a 100$\times$100 lattice, $t=-U/8$ and
$T=0.001U$ (the dashed lines). (a) $\mu=0.5U$, $\eta=0.01U$. (b)
$\mu=0.1U$, $\eta=0.02U$. The solid lines show the corrected $\Im
K(\omega)$ (see text).} \label{Fig_ii}
\end{figure}

Figure~\ref{Fig_ii} demonstrates $\Im K(\omega)$ calculated with the
use of equations~(\ref{approx}), (\ref{simplified})
and~(\ref{HubbardI}). The change to real frequencies carried out in the
previous section converts the Matsubara function~(\ref{GF}) into the
retarded Green's function \cite{Abrikosov}. It is an analytic function
in the upper half-plane which requires that $\Im K(\omega)$ be
negative. As seen from figure~\ref{Fig_ii},
this condition is violated at $\omega_d=-\mu$ and $U-\mu$. This
difficulty of the considered approximation was indicated in
reference~\cite{Pairault}. The problem is connected with divergencies
at these frequencies introduced by functions $g_{0\sigma}(\omega)$ and
$g_{\sigma 2}(\omega)$ in the above formulas. As can be seen from the
procedure of calculating the cumulants in the previous section, these
functions and divergencies with sign-changing residues appear in all
orders of the perturbation expansion~(\ref{series}). In the entire
series the divergencies are expected to compensate each other so that
the resulting $\Im K(\omega)$ is negative everywhere. However, in the
considered subset of terms such compensation does not occur.
Nevertheless, as seen from figure~\ref{Fig_ii}, at frequencies
neighboring to $\omega_d$ cumulants are regular and, if the used subset
of diagrams is expected to give a correct estimate of the entire series
for these frequencies, the value of $\Im K(\omega)$ at the singular
frequencies can be reconstructed using an interpolation and its values
in the regular region. Examples of such interpolation are given in
figure~\ref{Fig_ii}.

As seen from figure~\ref{Fig_ii}a, at half-filling, $\mu=\frac{U}{2}$,
$\Im K(\omega)$ has two broad minima. With decreasing the chemical
potential from this value these minima shift with respect to the Fermi
level without a noticeable change of their shapes until the Fermi level
enters one of the minima at $\mu\approx 0.17U$. As this takes place,
two new sharp minima arise near frequencies $-\mu$ and $U-\mu$ on the
background of the above-mentioned broad minima. The appearance of the
broad features in figure~\ref{Fig_ii} is connected with the third term
in $\sum_{\sigma_1}K_2$ in equation~(\ref{simplified}), while the sharp
minima are related to the second term in this formula. Its contribution
to $K(\omega)$, equation~(\ref{approx}), grows rapidly when the Fermi
level enters the broad minimum.

The function $K(z)$ has to be analytic in the upper half-plane also and
therefore its real part can be calculated from its imaginary part using
the Kramers-Kronig relations. We use this way with the interpolated
$\Im K(\omega)$ to avoid the influence of the divergencies on $\Re
K(\omega)$. However, the use of the interpolation overrates somewhat
values of $|\Im K(\omega)|$ which leads to the overestimation of the
tails in the real part. To correct this defect the interpolated
$K(\omega)$ is scaled so that in the far tails its real part coincides
with the values obtained from equation~(\ref{approx}).

\begin{figure}
\centerline{\includegraphics[width=0.46\textwidth]{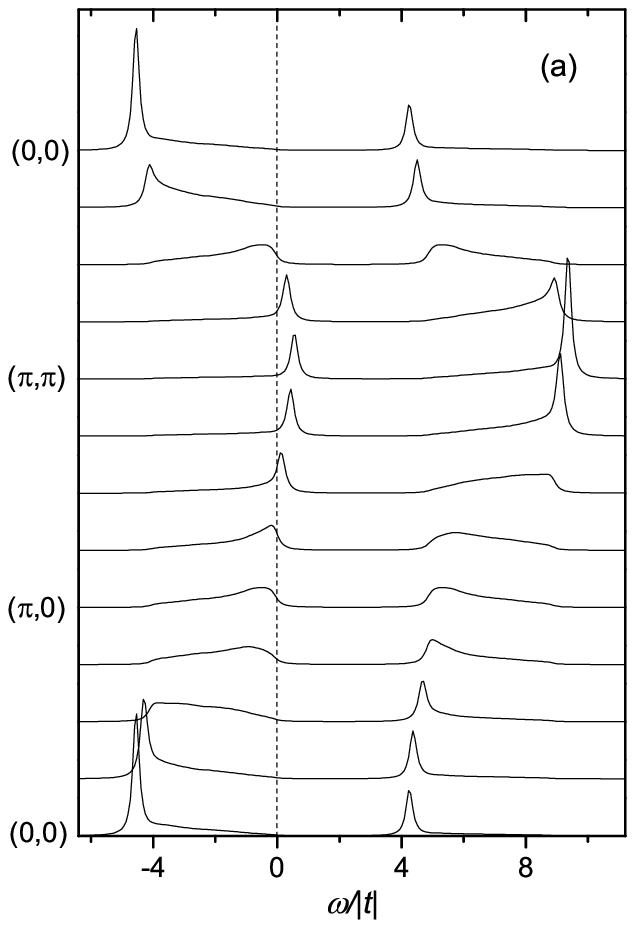}\hspace{2em}
\raisebox{-.5ex}{\includegraphics[width=0.46\textwidth]{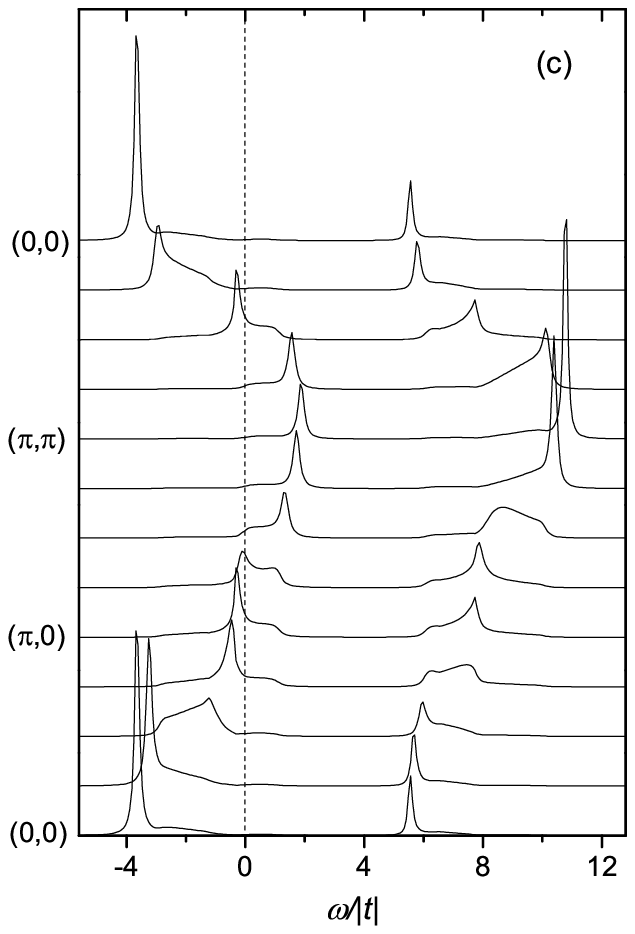}}}

\vspace{3ex}
\hspace*{.5em}\centerline{\includegraphics[width=0.48\textwidth]{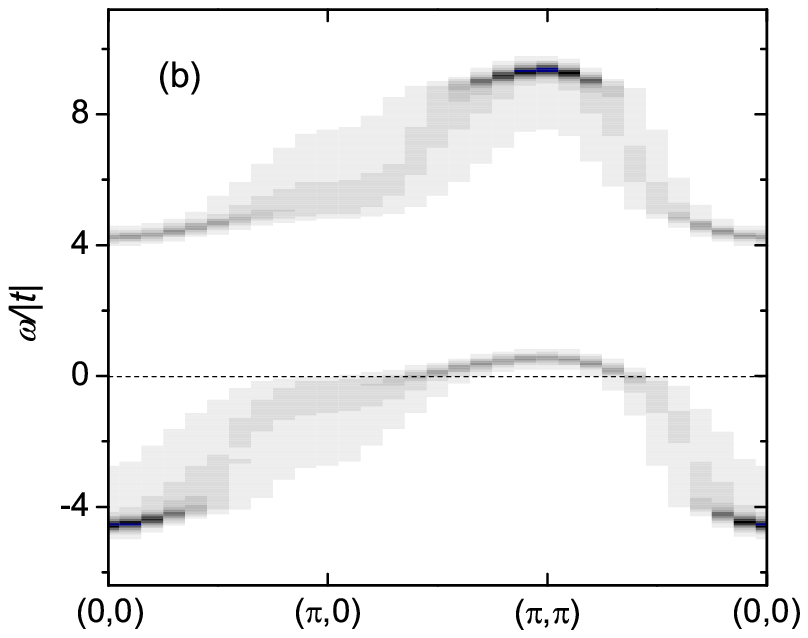}
\hspace{.5em}\includegraphics[width=0.48\textwidth]{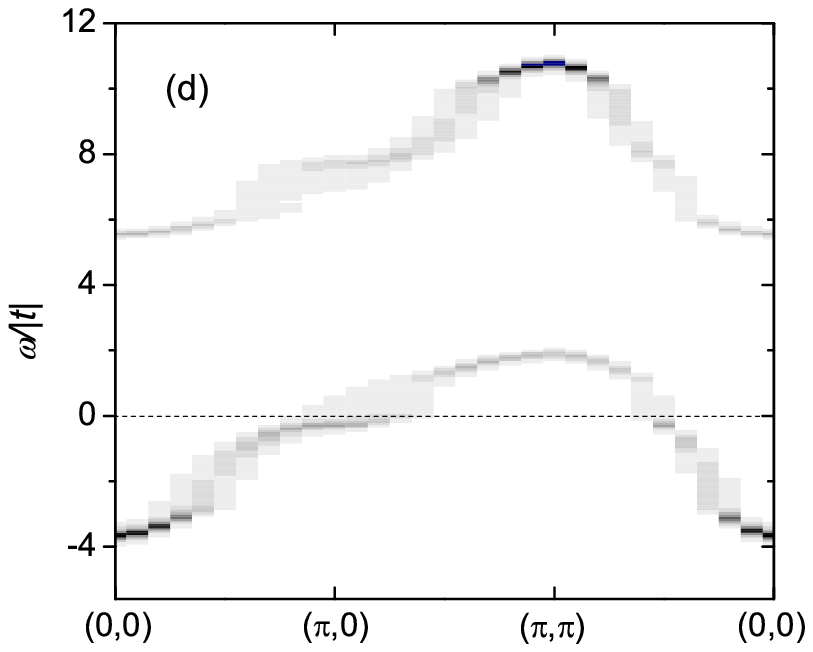}}
\caption{Panels (a) and (c): the spectral function $A({\bf k}\omega)$
calculated for momenta along the symmetry lines of the square Brillouin
zone in a 100$\times$100 lattice for $t=-U/8$, $T=0.001U$,
$\eta=0.02U$, $\mu=0.2U$ (a) and $\mu=0.05U$ (c). The dispersions of
maxima in the panels (a) and (c) are shown in the panels (b) and (d),
respectively. Here darker areas correspond to larger intensities of
maxima.} \label{Fig_iii}
\end{figure}

The spectral function
\begin{equation}
A({\bf k}\omega)=-\frac{1}{\pi}\Im G({\bf k}\omega)=-\frac{1}{\pi}
\frac{\Im K(\omega)}{[1-t_{\bf k}\Re
 K(\omega)]^2+[t_{\bf k}\Im K(\omega)]^2}\label{specfun}
\end{equation}
obtained in this way for momenta along the symmetry lines of the square
Brillouin zone is shown in figure~\ref{Fig_iii}.
The shapes of the spectral function in figure~\ref{Fig_iii}a are nearly
the same as at half-filling -- with decreasing $\mu$ from $0.5U$ to
approximately $0.17U$ these curves shift with respect to the Fermi
level without perceptible changes in their shapes. As indicated in
reference~\cite{Sherman}, four bands can be distinguished in these
spectra. For parameters of figure~\ref{Fig_iii}a these bands are
located near frequencies $-4|t|$, $|t|$, $4|t|$, and $9|t|$ (see
figure~\ref{Fig_iii}b). For the major part of the Brillouin zone the
peaks forming the bands arise at frequencies which satisfy the equation
$1-t_{\bf k}\Re K(\omega)=0$ and fall into the region of a small
damping $|\Im K(\omega)|$ [see equation~(\ref{specfun})]. As seen from
figure~\ref{Fig_ii}, such regions of small damping are located between
and on the outside of the two broad minima in $\Im K(\omega)$. This is
the reason of the existence of the four well separated bands -- two of
them are located between the minima of $\Im K(\omega)$, while two
others are on the outside of these minima. Broader maxima of $A({\bf
k}\omega)$ for momenta near the boundary of the magnetic Brillouin zone
are of different nature -- since $t_{\bf k}$ is small for such momenta,
the resonant denominator in equation~\ref{specfun} does not vanish and
the shape of the spectral functions is determined by $\Im K(\omega)$ in
the numerator of this formula.

More substantial changes in $A({\bf k}\omega)$ occur for $\mu\leq
0.17U$ when the Fermi level enters one of the broad minima in $\Im
K(\omega)$. As seen from figure~\ref{Fig_iii}c, in addition to the
mentioned four bands there appear sharp dispersive features near
$\omega=-\mu$ and $U-\mu$ for momenta in the vicinity of the boundary
of the magnetic Brillouin zone. It is clear that these changes in the
spectral function are connected with the changes in $\Im K(\omega)$
shown in figure~\ref{Fig_ii}. The peaks near $-\mu$ are more intensive
than those near $U-\mu$ and are located in the nearest vicinity of the
Fermi level. For $\mu\approx 0.17U$ the width of the band formed by the
former peaks is comparable to the superexchange constant $J=4t^2/U$ of
the effective Heisenberg model which describes magnetic excitations in
the limit $U\gg|t|$. This indicates obviously the participation of the
spin excitations in the formation of these band states. The bandwidth
decreases with further reduction of the electron concentration. As this
takes place, the peak intensities first grow and then saturate. The
maximum energies of the band are located near the boundary of the
magnetic Brillouin zone. For parameters of figure~\ref{Fig_iii}c for
these momenta the band touches the Fermi level and the corresponding
peaks disappear above it (see figure~\ref{Fig_iii}d).

These properties of the band resembles those of the spin-polaron band
in the $t$-$J$ model. This latter band is also located near the Fermi
level, has the similar dispersion and the bandwidth, which decreases
with the reduction of the electron concentration (with the increase of
the hole doping counted from half-filling) \cite{Sherman94}. However,
there is one essential difference in the behavior of these bands in the
Hubbard and $t$-$J$ models. In the former model the narrow band near
the Fermi level appears at a certain deficiency of electrons, while the
above-mentioned four bands exist in the entire considered range of
electron concentrations. In the $t$-$J$ model the situation is opposite
-- the spin-polaron band exist in the wide range of hole concentrations
$0\leq x \leq 0.17$, while the wider band -- an analog of the four-band
structure -- starts to form at $x\approx 0.06$ \cite{Sherman94}.

The value of the chemical potential for which the Fermi level enters
the minimum of $\Im K(\omega)$ and the narrow band begins to form
depends on the ratio $t/U$. For example, for $t=-U/4$ this happens at
$\mu\approx 0.27U$.

There is some difficulty in the comparison of our results with the data
of Monte-Carlo calculations and cluster theories. It is connected with
the fact mentioned in reference~\cite{Sherman}: the one-loop
approximation overestimates the spectral weights of the two internal
bands of the four-band structure near the momenta $(0,0)$ and
$(\pi,\pi)$. As a consequence the electron concentration calculated
from the formula
\[\langle n\rangle=\frac{2}{N}\sum_{\bf
k}\int_{-\infty}^\infty{\rm d}\omega\frac{A({\bf
k}\omega)}{\exp(\beta\omega)+1}\] appears to be considerably
underestimated. For example, for the parameters of
figures~\ref{Fig_iii}a and~\ref{Fig_iii}c such estimated $\langle n
\rangle$ equals to $0.87$ and $0.68$, respectively. From the comparison
of the dispersions in figures~\ref{Fig_iii}b and~\ref{Fig_iii}d with
the results of Monte-Carlo calculations (figure~9 in \cite{Grober}) it
can be concluded that these concentrations have to be approximately
$0.95$ and $0.85$, respectively. With this in mind we find that the
spectral functions and dispersions in figure~\ref{Fig_iii} are close to
the Monte-Carlo spectra (cf.\ with figures~10 and~11 in \cite{Grober}).
In these latter spectra for some deviation from half-filling a weakly
dispersive feature is also observed near the Fermi level. However, this
feature has low intensity and is lost at the foot of a more intensive
maximum on approaching the boundary of the magnetic Brillouin zone.
These differences may be connected with the comparative high
temperature $T=0.33|t|$ used in the Monte-Carlo simulations. A similar
weakly dispersive feature near the Fermi level was obtained also in
quantum cluster theories, however for much smaller deviations from
half-filling (cf. with figure~2c in \cite{Kyung} and figure~32 in
\cite{Maier}).

\section{Conclusion}
The considered diagram technique is very promising for a generalization
to many-band Hubbard models for which energy parameters of the one-site
parts of the Hamiltonians exceed or at least are comparable to the
intersite parameters. The expansion in powers of these latter
parameters can be expressed in terms of cumulants in the same manner as
discussed above. Now there are distinct cumulants which belong to
different site states characterized by dissimilar parameters of the
repulsion and the level energy. These cumulants are described by
formulas similar to equations~(\ref{firstc}) and~(\ref{secondc}). For
example, for the Emery model \cite{Emery} which describes oxygen
$2p_\sigma$ and copper $3d_{x^2-y^2}$ orbitals of Cu-O planes in
high-$T_c$ superconductors there are two types of cumulants
corresponding to these states. Diagrams of the lowest orders, e.g., for
Green's function on copper sites resemble those shown in
figure~\ref{Fig_i} where ``oxygen'' cumulants are included in hopping
lines. Equations of the type of (\ref{tLarkin}) and partial summations
similar to equation~(\ref{thopping}) can be derived also in this case.

In summary, the diagram technique for the one-band Hubbard model was
formulated for the case of moderate to strong Hubbard repulsion. The
expansion in powers of the hopping constant is expressed in terms of
site cumulants of electron creation and annihilation operators. For
Green's function the equation of the Larkin type was derived and solved
for the case of two dimensions and nearest-neighbor hopping. With
decreasing the electron concentration in addition to the four bands
observed at half-filling a narrow band arises near the Fermi level for
momenta near the boundary of the magnetic Brillouin zone. On the
occurrence the width of the band is comparable to the superexchange
constant $J=4t^2/U$ which indicates the participation of the spin
excitations in the band formation. The bandwidth decreases with
decreasing the electron concentration. The maximum energies of the band
are located near the boundary of the magnetic Brillouin zone. For some
deviation from half-filling in these points the band touch the Fermi
level. By these properties the band resembles the spin-polaron band of
the $t$-$J$ model.

\end{document}